\documentclass[a4paper,
               biblatex     
               ]{jacow}
\usepackage{pdfpages,multirow,ragged2e} %
\makeatletter%
	\ifboolexpr{bool{xetex}}
	 {\renewcommand{\Gin@extensions}{.pdf,%
	                    .png,.jpg,.bmp,.pict,.tif,.psd,.mac,.sga,.tga,.gif,%
	                    .eps,.ps,%
	                    }}{}
\makeatother

\ifboolexpr{bool{xetex} or bool{luatex}} 
 {}                                      
 {\usepackage[utf8]{inputenc}}           

\usepackage[USenglish]{babel}

\ifboolexpr{bool{jacowbiblatex}}%
 {%
  \addbibresource{THIAA04.bib}
 }{}
\begin{document}

\title{PERFORMANCE ANALYSIS OF SPOKE RESONATORS, STATISTICS FROM CAVITY FABRICATION TO CRYOMODULE Testing}

\author{A.~ Miyazaki\thanks{akira.miyazaki@ijclab.in2p3.fr}, P.~Duchesne, D.~Ledrean, D.~Longuevergne, and G.~Olry, \\ CNRS/IN2P3/IJCLab Universit\'{e} Paris-Saclay, 91405 Orsay, France}
	
\maketitle

\begin{abstract}
Ir\`{e}ne Joliot-Curie Laboratory (IJCLab) has been leading the development of spoke resonators in multiple international SRF projects, from fundamental R\&D, prototyping, to series production. 
The European Spallation Source (ESS) superconducting linac is the first of its kind to put into operation the spoke resonators. 
After three prototype cavities, 29 ESS production cavities have been processed, tested, assembled into cryomodules at IJCLab, and then shipped to Uppsala for the site acceptance test. 
Seven prototypes for two other major projects, Multi-purpose hYbrid Research Reactor for High-tech Application (MYRRHA) and Proton Improvement Plan II (PIP-II), designed in collaboration with external institutions, have as well been processed and tested at IJCLab. 
A new challenge is to fully process series cavities in industry, following the successful implementation of 1.3~GHz elliptical cavities in the other projects. 
This paper summarises main results obtained from fabrication to final testing, including frequency tuning strategy, performance, limitation in vertical cryostat, and identifies future direction of projects and R\&D in the field of spoke cavities.
\end{abstract}

\section{INTRODUCTION}
Superconducting spoke cavities are the choice in a medium-$\beta$ section of proton drivers.
Since the late 1980s~\cite{DELAYEN1989892}, spoke cavities have been developed and their technology is matured today~\cite{PhysRevSTAB.6.080101, SHEPARD2006205, OLRY2006201, PhysRevSTAB.16.102001, Li_2014}.
However, practical challenges for the deployment of the these cavities in real machines need to be identified and overcome.
Unlike the 1.3~GHz TESLA-type cavities, spoke cavities are not sufficiently standardised and there are many open questions towards the successful operation of accelerators.
Ir\`{e}ne Joliot-Curie Laboratory (IJCLab) plays a leading role in this crucial matter in international projects.

IJCLab has been pioneering the development of spoke cavities technology from fundamental studies~\cite{PhysRevAccelBeams.24.083101, LONGUEVERGNE201841} to design and prototyping work~\cite{OLRY2006201, DUTHIL2016} and even deployment in the machines.
In this paper, we overview state-of-the-art technology in developing spoke resonators at IJCLab with three international projects as examples.
The series production of European Spallation Source (ESS)~\cite{Peggs:2013sgv} double-spoke cavities revealed delicate issues in frequency tuning, including fabrication, chemical etching, and heat treatment.
We discuss how we overcame these issues with statistics obtained during the production.
These ESS cavities have been qualified in cold tests, integrated in cryomodules, and all passed the site-acceptance tests at Uppsala University.
The next challenge of ESS is installation, commissioning, and of course, operation in the machine.

We completed prototyping four single-spoke cavities for the Multi-purpose hYbrid Research Reactor for High-tech Application (MYRRHA)~\cite{0976c72ed89a41e6a8e2e1eb936af11c}.
The next challenge is to industrialise the surface treatment of spoke cavities, 
whose complicated shape may require special attentions compared to conventional elliptical cavities.
We also show preliminary results on heat treatment in prototype MYRRHA cavities, which may be a breakthrough towards 4~K operation of spoke resonators.
Finally, we started prototyping Single Spoke Resonator 2 (SSR2) for Proton Improvement Plan II (PIP-II).
We discuss preliminary results and trade-off of cavity design between RF performance and cleaning process.

\section{SUPRATECH}
\begin{figure}[!htb]
   \centering
   \includegraphics*[width=65mm]{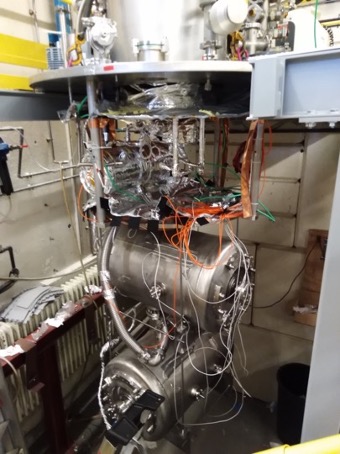}
   \caption{Vacuum insert of the vertical cryostat at SUPRATECH. Two cavities are mounted in the same insert and share the same beam vacuum.}
   \label{fig:C800}
\end{figure}

IJCLab hosts the SUPRATECH facilities, where one can perform chemical treatments, high-pressure water rinsing (HPR), heat treatment, mechanical frequency tuning, cold tests at 4~K and 2~K, assembly of cavity string and cryomodules, and cryomodule testing.
As shown in Fig.~\ref{fig:C800}, two spoke cavities can be tested at a time in a vertical cryostat ($\phi800$) with a vacuum insert which requires a helium tank welded around the cavity~\cite{PhysRevAccelBeams.24.083101}.
In SUPRATECH, we do not measure bare cavities today because of the following strategic reasons.
First, we can save a huge amount of liquid helium for cold tests, because helium supply is a global issue for the SRF community today.
Secondly, the small heat capacitance of the cryostat enables quick cool down and warm up.
As a drawback, however, careful frequency tuning is required in fabrication, surface processing, and even cooling down because we skip the cold test of a bare cavity to check the frequency before welding the helium tank.
This unique tuning and testing strategy has been successful in the ESS project and the same was adopted for other similar projects (MYRRHA and PIP-II SSR2).
Therefore, a global standard of future projects can follow our strategy: accurate frequency tuning and only one cold test in a vertical test-stand.

\section{ESS SERIES CAVITIES}
ESS is a proton driver in Sweden for neutron science via nuclear spallation.
In the spoke section from 90~MeV to 216~MeV,
it deploys 13 cryomodules, each of which accommodates two double-spoke cavities with 352~MHz.
The target gradient is $E_{\rm acc}=9$~MV/m with an unloaded quality factor $Q_0=1.5\times10^{9}$ at 2~K.
Since the geometry factor is $G=133$~$\Omega$ and the peak field ratio is $6.9\, {\rm mT(MV/m)^{-1}}$, 
the target surface resistance and peak magnetic field are 89~n$\Omega$ and 69~mT, respectively.
Compared to state-of-the-art elliptical cavities, 
the target performance is more conservative partially because the field level is somewhat limited by beam dynamics in the spoke section.
However, we achieved performance beyond the specification by one order of magnitude.
Typical $Q_0$ achieved was above $2\times10^{10}$ at 9~MV/m and maximum gradient was above 15~MV/m.

A particular challenge of these cavities was frequency tuning without a bare cavity testing at cold.
The behavior of spoke cavities, with intrinsically complicated shape compared to the conventional elliptical cavities, must be fully understood and controlled in such a very delicate process.
We describe the strategy in fabrication, heat treatment, and chemical etching in the following subsections.

\subsection{TUNING AT FABRICATION}
\begin{figure}[!htb]
   \centering
   \includegraphics*[width=45mm]{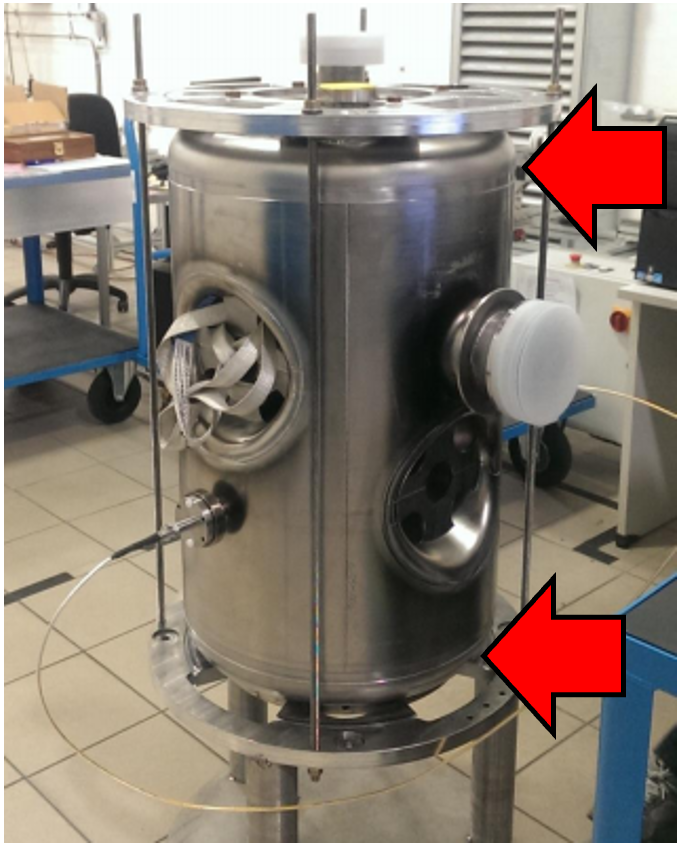}
   \caption{Configuration of the frequency measurement before the trimming and final electron-beam welding. The red arrows indicate the trimming margins and the end-caps are temporarily clamped for the RF measurement at warm.}
   \label{fig:trimming}
\end{figure}

The goal of fabrication is to keep the frequency tolerance within $\pm150$~kHz.
The body of the cavity has 5~mm margin for both sides before welding the end-caps there, as indicated by red arrows in Fig.~\ref{fig:trimming}.
Depending on the frequency, preliminary measured by clamping these parts, and based on frequency shifts by the electron-beam welding (EBW), the trimming lengths were decided by IJCLab.
The first leak check was performed after this final EBW and frequency was permanently shifted due to vacuum pumping inside of the cavities.

\begin{figure}[!htb]
   \centering
   \includegraphics*[width=85mm]{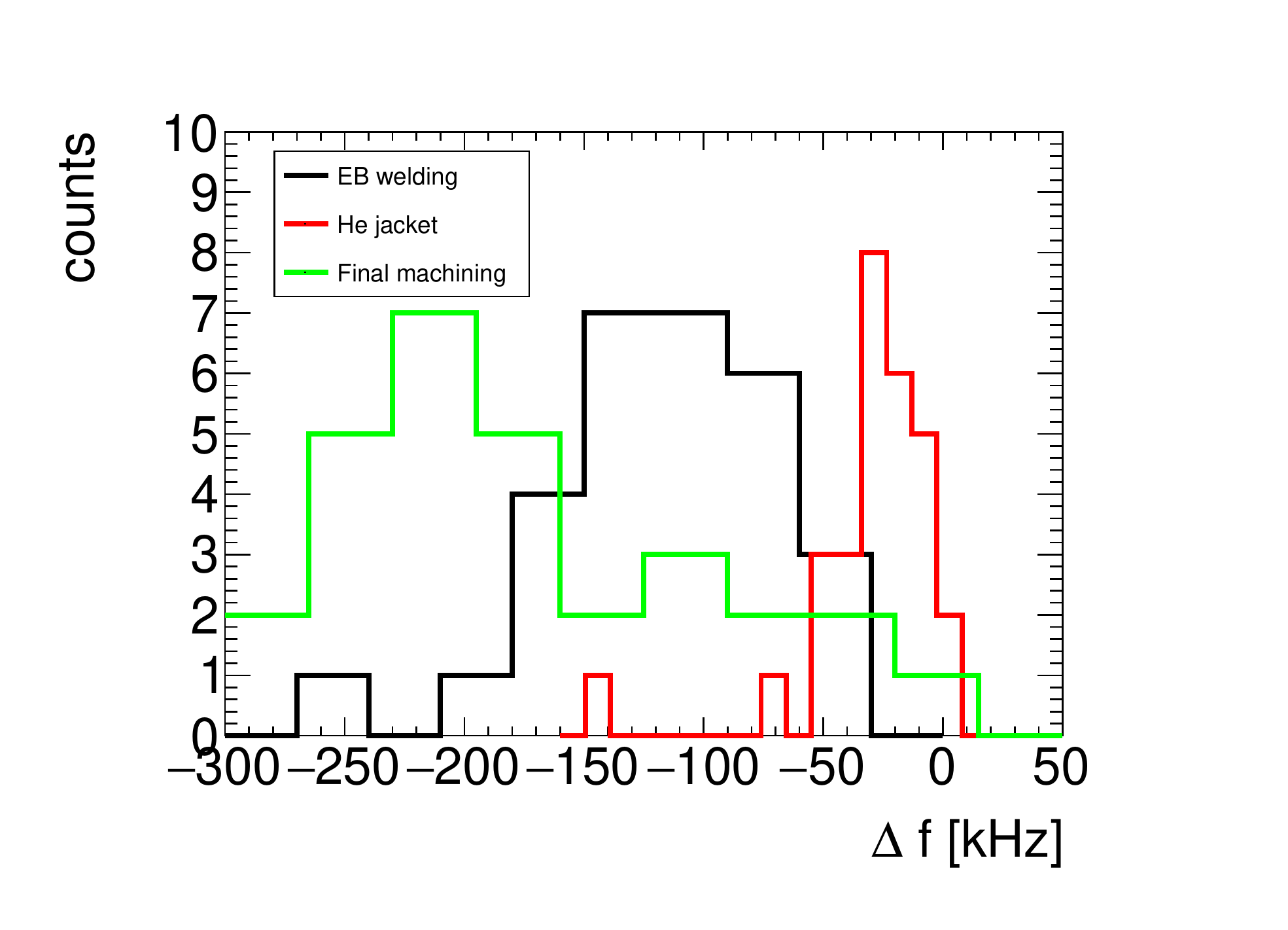}
   \caption{Statistical distributions of frequency shifts due to the final EB welding, TIG welding of the helium jacket, and final machining.}
   \label{fig:tune_machine}
\end{figure}

After the first leak check, a helium jacket was manually welded to the cavity body by Gas tungsten arc (TIG) welding.
In the beginning of the series production, frequency shifts by different welders were evaluated by four pre-series cavities in order to estimate the influence of this manual welding on the helium jacket.
The jacket welding of series cavities were performed by the selected welder and further statistics were recorded.
After the welding, careful machining was performed in order to form the parts, which are dedicated to mounting cold tuners on the cavity during the module assembly process.
The cavities frequency was further shifted due to a strong supporting force that holds the cavity during machining because of the required high mechanical tolerance for the tuner.
As summarized in Fig.~\ref{fig:tune_machine}, although a few exceptions were observed,
the frequency tuning during the manufacturing process was under control.

\subsection{CHEMICAL TREATMENT}
\begin{figure}[!htb]
   \centering
   \includegraphics*[width=85mm]{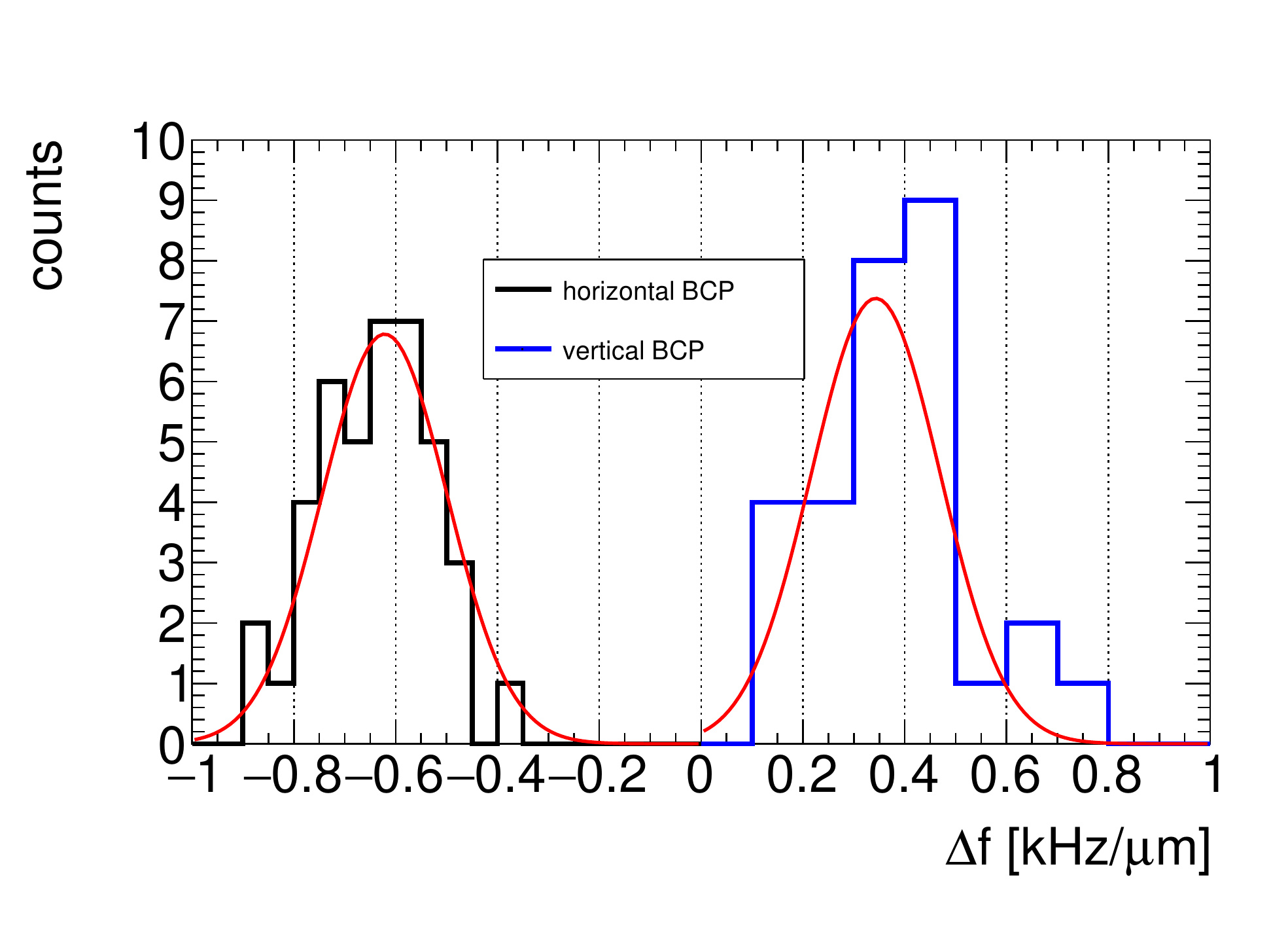}
   \caption{Statistics of frequency detuning by BCP. The Gaussian fits show mean values of $-0.62\, {\rm kHz/\mu m}$ and $+0.34\, {\rm kHz/\mu m}$ for horizontal and vertical BCPs, respectively.}
   \label{fig:BCP}
\end{figure}

The chemical treatment is the major mean to fine-tune the cavity frequency after the manufacturing process.
Ports originally prepared for HPR,
enabled Buffer Chemical Polishing (BCP) in two orientations.
Horizontal BCP decreases the frequency by $-0.62$~kHz/$\mu$m while vertical one increases it by $+0.34$~kHz/$\mu$m as shown in Fig.~\ref{fig:BCP}.
Depending on the frequency at the reception, 
we optimised the periods of horizontal and vertical BCPs,
in order to meet the frequency tolerance within $\pm40$~kHz.
After heat treatment,
light BCP was performed to remove surface contamination generated during annealing.
This light BCP is mainly in the vertical orientation;
however, fine-tuning by additional horizontal BCP was sometimes necessary due to unexpected frequency detuning caused by the heat treatment as described in the next subsection.

\subsection{HEAT TREATMENT}
After bulk BCP, heat treatment was performed in order to degas hydrogen and avoid Q-disease.
The annealing parameters were optimised to 650$^\circ$C for 10 hours because higher temperature was not necessary due to little gain by flux expulsion as described in Ref.~\cite{PhysRevAccelBeams.24.083101}.
Since the annealing temperature is marginal, some flanges were even copper-brazed in advance.

\begin{figure}[!htb]
   \centering
   \includegraphics*[width=85mm]{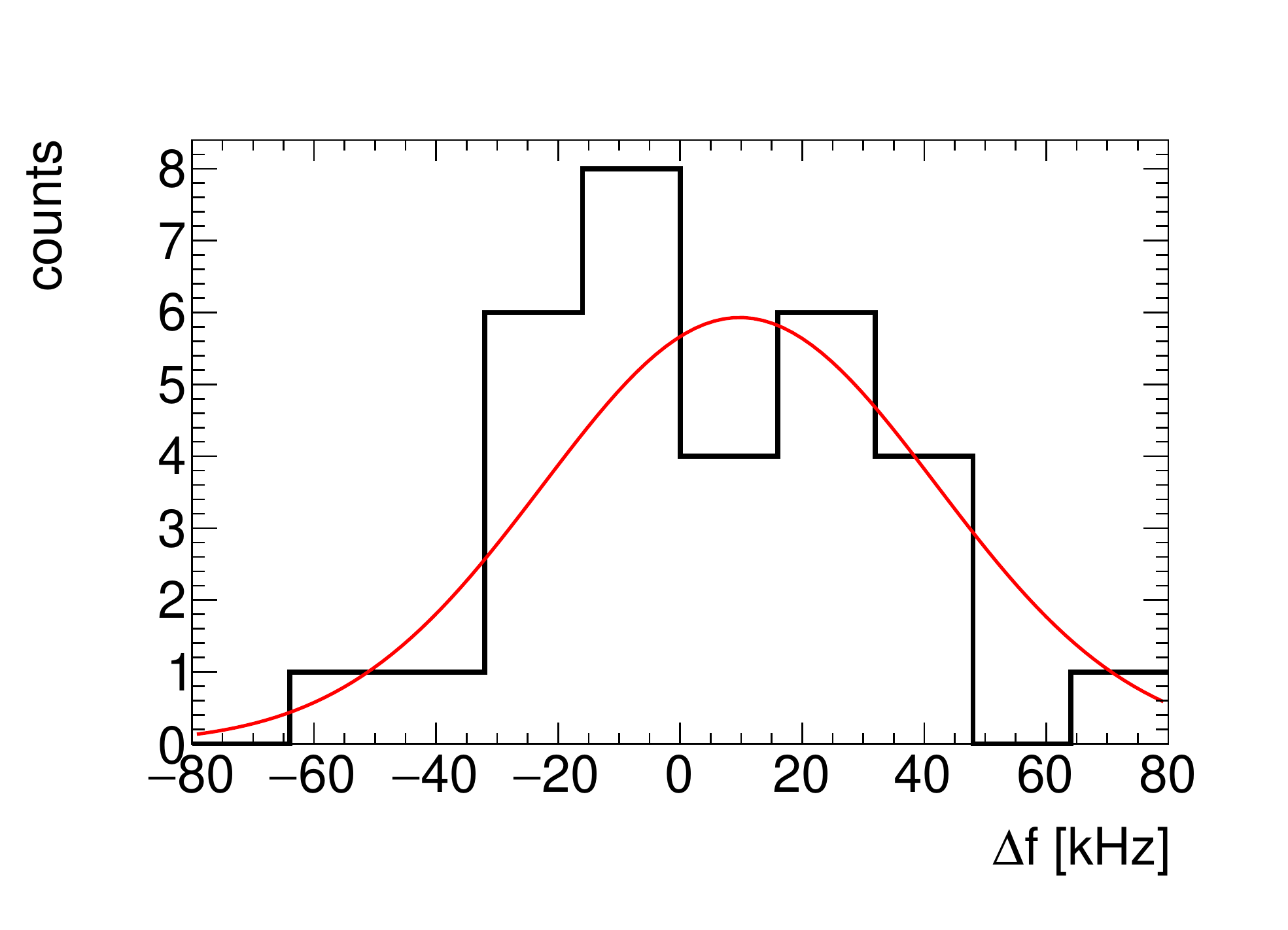}
   \caption{Statistics of frequency detuning by annealing at 650$^\circ$C for 10~h. The Gaussian fit shows the mean at $9.8\, {\rm kHz}$ with standard deviation of $32\, {\rm kHz}$.}
   \label{fig:four}
\end{figure}

The frequency shift by heat treatment showed an unexpected behavior as seen in Fig.~\ref{fig:four}.
It statistically distributes around $+10$~kHz with a substantially large standard deviation at $32$~kHz.
The heat treatment either increases or decreases the resonant frequency of cavities unpredictably.
This may be due to the helium jacket (made of titanium) annealed together with the niobium cavity, releasing mechanical stress in its material history.
We did not know the stress level of this titanium jacket at the stage of heat treatment.
In the series production of the ESS cavities,
we made use of horizontal and vertical BCPs to compensate the unexpected frequency shift.
In a few cases, we performed mechanical tuning by pressurising the helium circuit.
For more details, see Ref.~\cite{duchesne:ipac2021-mopab392}.

\subsection{COLD TESTS}
\begin{figure}[!htb]
   \centering
   \includegraphics*[width=85mm]{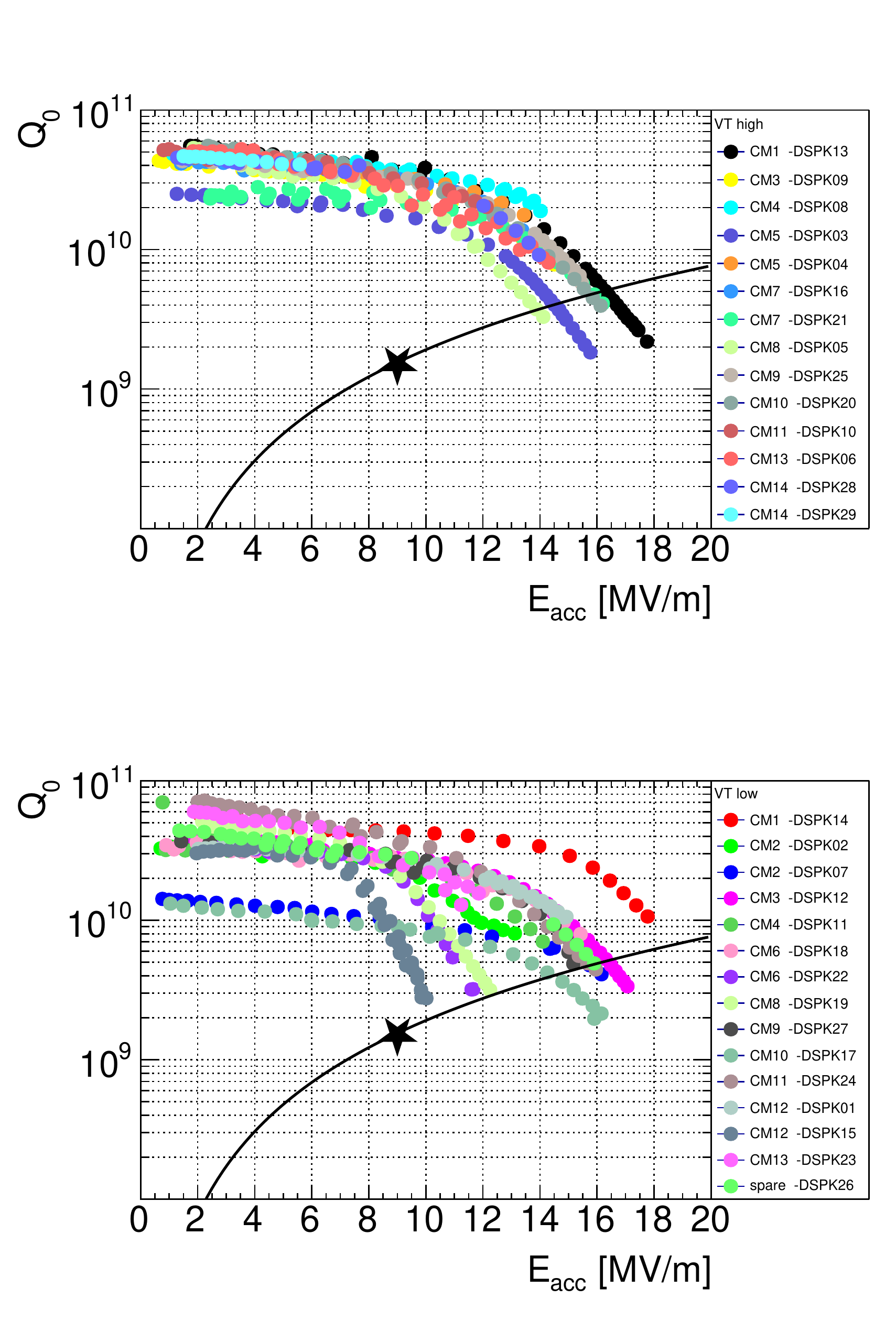}
   \caption{RF performance of the ESS series cavities mounted at higher (upper figure) and lower (lower figure) position in the vacuum insert.}
   \label{fig:ess_Q_vs_E}
\end{figure}

In order to qualify the RF performance of the cavities, cold tests were performed after HPR.
All 29 cavities passed the tests although a few of them required several iterations with HPR and sometimes even light BCP to remove contaminants causing field emission.
We obtained excellent performance in all the cavities as shown in Fig.~\ref{fig:ess_Q_vs_E}, with low-field surface resistance ranging from 2 to 7~n$\Omega$.
In order to evaluate the trapped flux sensitivity for the series cavities, 
two cavities (DSPK07 and 17) were measured without active compensation of the ambient magnetic field and still met the project specification.
This is consistent to the dedicated studies with prototype cavities~\cite{PhysRevAccelBeams.24.083101}.
Another concern was the fact that the cavity mounted at the lower side might capture more contamination than the upper one, because the lower one got cold even when the upper was still at room temperature.
Nevertheless, as shown in Fig.~\ref{fig:ess_Q_vs_E}, we did not find significant systematic differences in the cavities tested at upper or lower positions of the vacuum insert.

Note that no HPR was performed between the cold test and module assembly\footnote{HPR was performed for two cavities DSPK06 and 13 exceptionally after the cold test and field emission in DSPK23 observed in the test was successfully removed in the cryomodule. Although the pick-up antennas were carefully re-mounted after the HPR, the field calibration could be potentially uncertain. However, no major impact was observed in the site acceptance test.}.
Therefore, the pick-up antenna was not touched until the site acceptance tests so that the calibrated field to power values were preserved.
This may provoke a concern about degradation of field emission onset in the cryomodule,
but we did not observe any substantial increase of X-rays in the site acceptance test at Uppsala University~\cite{FREIA_SRF2023_poster}.
This evidences that ICJLab's strategy was successful.

\section{MYRRHA PROTOTYPE CAVITIES}
A proton driver in MYRRHA will provide a high-power proton beam to an accelerator-driven subcritical nuclear reactor.
The first stage of the accelerator is composed of 60 single spoke cavities to provide protons at 100~MeV.
IJCLab developed four prototype single spoke cavities in collaboration with SCK-CEN.
The cavities will be operated at 352~MHz and its target gradient is at 9~MV/m including fault tolerance during the reactor operation.
The peak field ratio of MYRRHA's single spoke cavities is slightly higher than that of ESS, $7.3\, {\rm mT/(MV/m)^{-1}}$ while the geometrical factor is slightly lower $G=109\, {\rm \Omega}$.

\begin{figure}[!htb]
   \centering
   \includegraphics*[width=85mm]{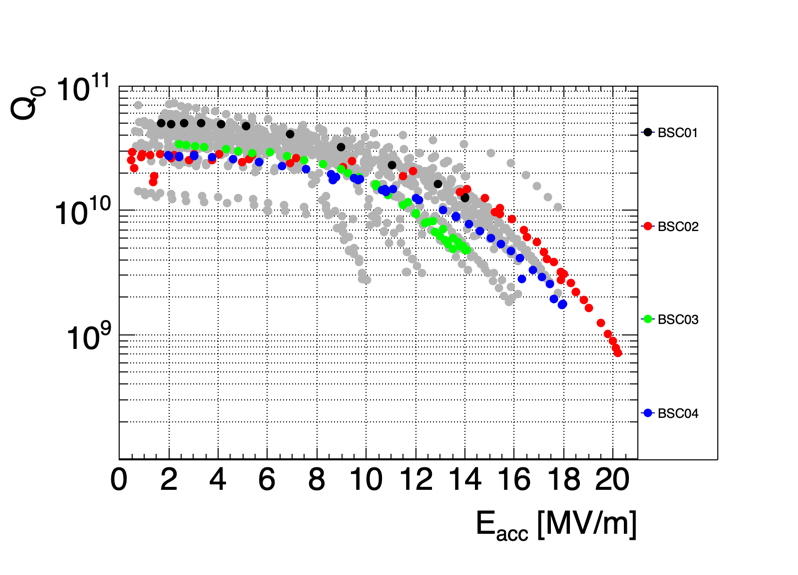}
   \caption{RF performance of MYRRHA prototype cavities projected onto the ESS series cavities. The difference in the geometrical factors are disregarded in this comparison.}
   \label{fig:MYRRHA}
\end{figure}

The RF performance is shown in Fig.~\ref{fig:MYRRHA} plotted on top of all the ESS series cavity results in a gray scale.
MYRRHA prototype cavities showed as excellent performance as ESS and therefore the future series production looks promising.
Note that the peak magnetic field and surface resistance are slightly different in ESS and MYRRHA due to geometrical factors, but this does not have any impact in this conclusion.

The next challenge in the MYRRHA project is full industrialization of all the production process including chemical, heat treatments, and HPR.
This is one of the major objectives of the pre-series cavity development led by SCK-CEN.
IJCLab contributes to vertical tests of the pre-series cavities as well as giving practical advice based on our long experience in spoke cavity development for ESS.
Note that such industrialization has been successful in 1.3~GHz TESLA-type elliptical cavities but its application to spoke cavities is highly non-trivial due to its fundamentally complicated shape intrinsically required by the RF performance at low-$\beta$.

\section{BAKING OF SPOKE CAVITIES}
During the series production of ESS double-spoke cavities,
IJCLab did not perform conventional baking after HPR except for very mild heating (120$^\circ$C, 3~h) for just drying water from the surface and reducing multipacting (MP).
Baking at 120$^\circ$C for 48~h is known as low temperature (low-T) baking~\cite{10.1063/1.1767295} and can improve the accelerating gradient and $Q_0$ in elliptical cavities at 1.3~GHz.
The former is thanks to removing the high-field Q-slope so that the peak magnetic field exceeds 100~mT (around 25~MV/m for typical elliptical cavities).
The latter is mainly thanks to lowering loss contributions from thermally excited quasi-particles on {\it dirtier} surface, so-called BCS resistance $R_{\rm BCS}$.
Consequently, low-T baking has been included in the standard procedure of other projects such as the International Linear Collider.
However, as a byproduct, a temperature-insensitive component, so-called residual resistance $R_{\rm res}$ usually increases.
Because of this issue, the spoke cavities do not necessarily benefit from this standard process. 

The field levels of low-$\beta$ cavities are limited by beam dynamics even if the ultimate field is improved. 
Typically, for the spoke cavities, around 9~MV/m or maximum 12~MV/m is the field level required by the projects.
Clearly, one does not need to remove the high-field Q-slope with low-T baking.
Moreover, at 2~K, the low frequency (below 400 MHz) leads to $R_{\rm BCS} < 1\, {\rm n\Omega}$ at 2 K because $R_{\rm BCS}$ has approximately a parabolic dependence on the RF frequency. 
In this case, $R_{\rm res}$ dominates the loss, so that low-T baking may even deteriorate the unloaded quality factor at low field.

\begin{figure}[!htb]
   \centering
   \includegraphics*[width=85mm]{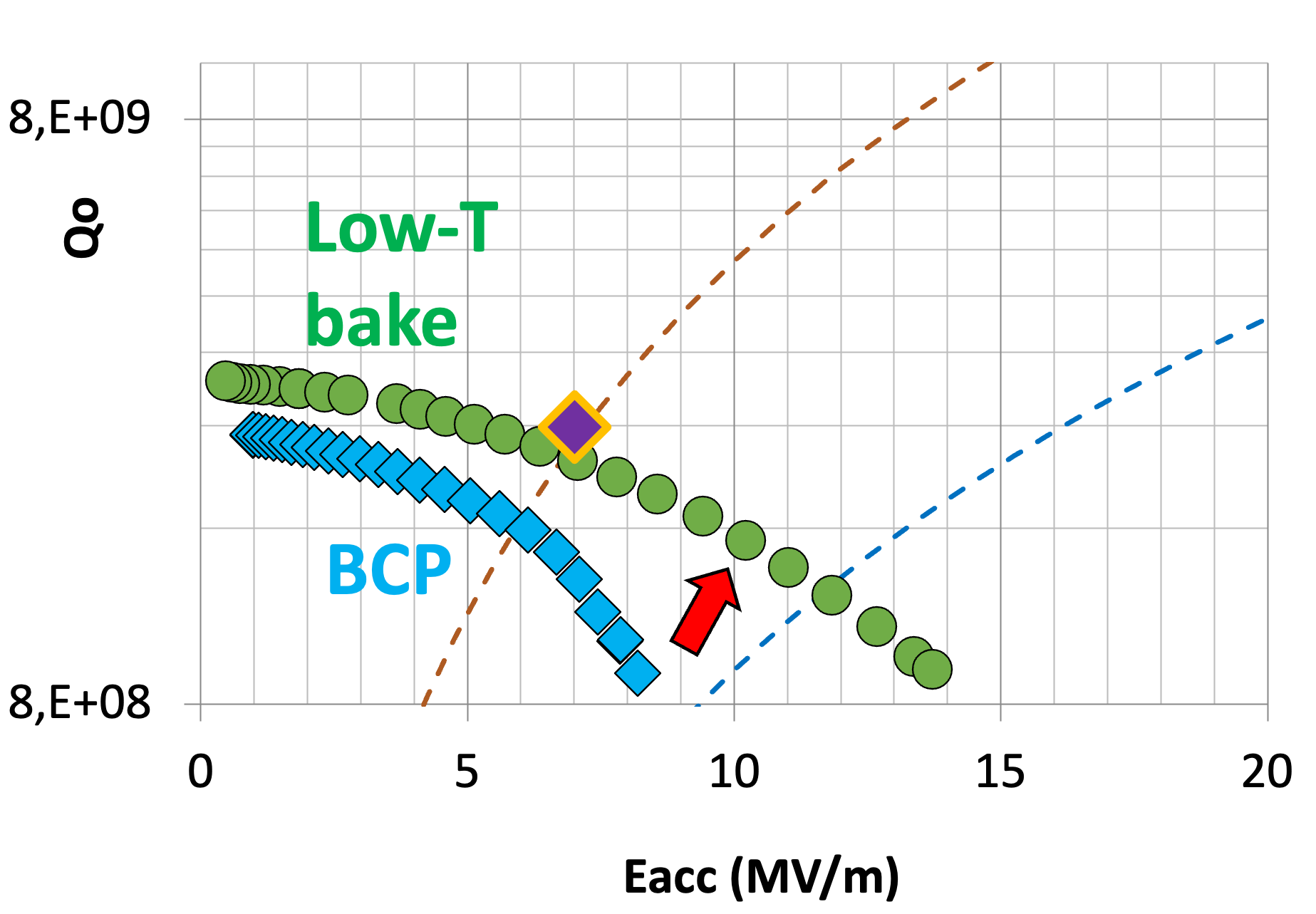}
   \caption{Influence of low temperature baking for a prototype MYRRHA cavity at 4~K.}
   \label{fig:MYRRHA_baking}
\end{figure}

When one takes into account the 4~K operation, 
the benefit of baking should be re-evaluated.
Since $R_{\rm BCS}$ is higher than $50\, {\rm n\Omega}$ at 4~K with 352~MHz,
the unloaded quality factor can be significantly improved by baking.
Figure.~\ref{fig:MYRRHA_baking} shows the substantial improvement of cavity performance at 4~K after low-T baking.
Although the MYRRHA project is primarily designed for the 2~K operation,
even the 4~K cavity performance after low-T baking met the specification of the machine.
This is a potential breakthrough in the future spoke cavity technology~\cite{Longuevergne:2018ykg}.

Considering the recent progress in various baking methods beyond the conventional low-T baking,
we suppose that the medium temperature baking (mid-T baking; 200-400$^\circ$C) may be the most promising as the new research direction.
IJCLab developed an excellent vacuum furnace~\cite{Fouaidy2021RecentRO} and we plan to perform fundamental studies in baking spoke cavities in coming years by using a spare ESS series cavity and prototype MYRRHA cavities.

\section{PIP-II PROTOTYPE CAVITIES}
The Proton Improvement Plan II (PIPII) is an international project to build a proton driver, hosted by Fermilab, for answering questions of fundamental physics, such as Dirac CP phase in neutrino, muon physics, and dark sector~\cite{Lebedev:2017vnu}.
This project includes two types of spokes cavities, SSR1 ($\beta=0.22$) and SSR2 ($\beta=0.47$), 
in which IJCLab is strongly involved on SSR2 section since the design phase~\cite{Parise:2019eod} and has agreed an in kind contribution for the production phase of SSR2~\cite{PIPII_SRF2023_poster}.

One objective of PIPII SSR2 is the same as MYRRHA, i.e., industrialization including fabrication and surface processing.
IJCLab performs the vertical tests of cavities fully prepared by the manufacturer and evaluate their quality of surface preparation.
When field emission is observed,
we perform our own surface treatment, fully qualified by ESS series cavities, and provide feedback to the manufacturer.

\begin{figure}[!htb]
   \centering
   \includegraphics*[width=90mm]{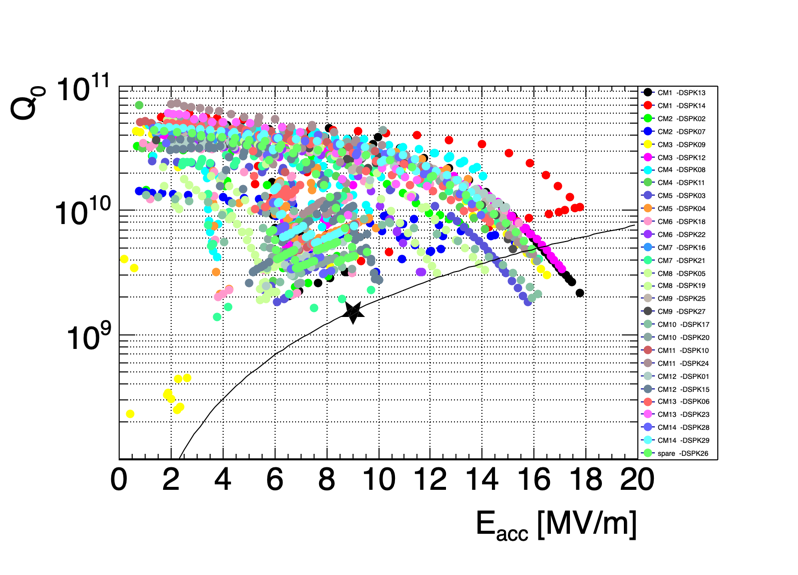}
   \caption{Q vs E curves of ESS cavities with multipacting bands.}
   \label{fig:ESS_MP}
\end{figure}

The design of SSR2 is based on lessons learned from SSR1 and ESS cavities.
As is well known, substantial MP is one of the major challenges of spoke cavities.
Figure.~\ref{fig:ESS_MP} shows the MP bands of ESS cavities before being conditioned.
Clearly, the MP bands even cover the nominal accelerating gradient, which is a typical field level for the spoke cavities.
This differentiates spoke cavities from the elliptical cavities whose MP bands are usually sufficiently lower than the operational gradient.
The potential concern is any unexpected influence in stability during accelerator operation even if the MP bands are conditioned in advance during machine commissioning.

\begin{figure}[!htb]
   \centering
   \includegraphics*[width=85mm]{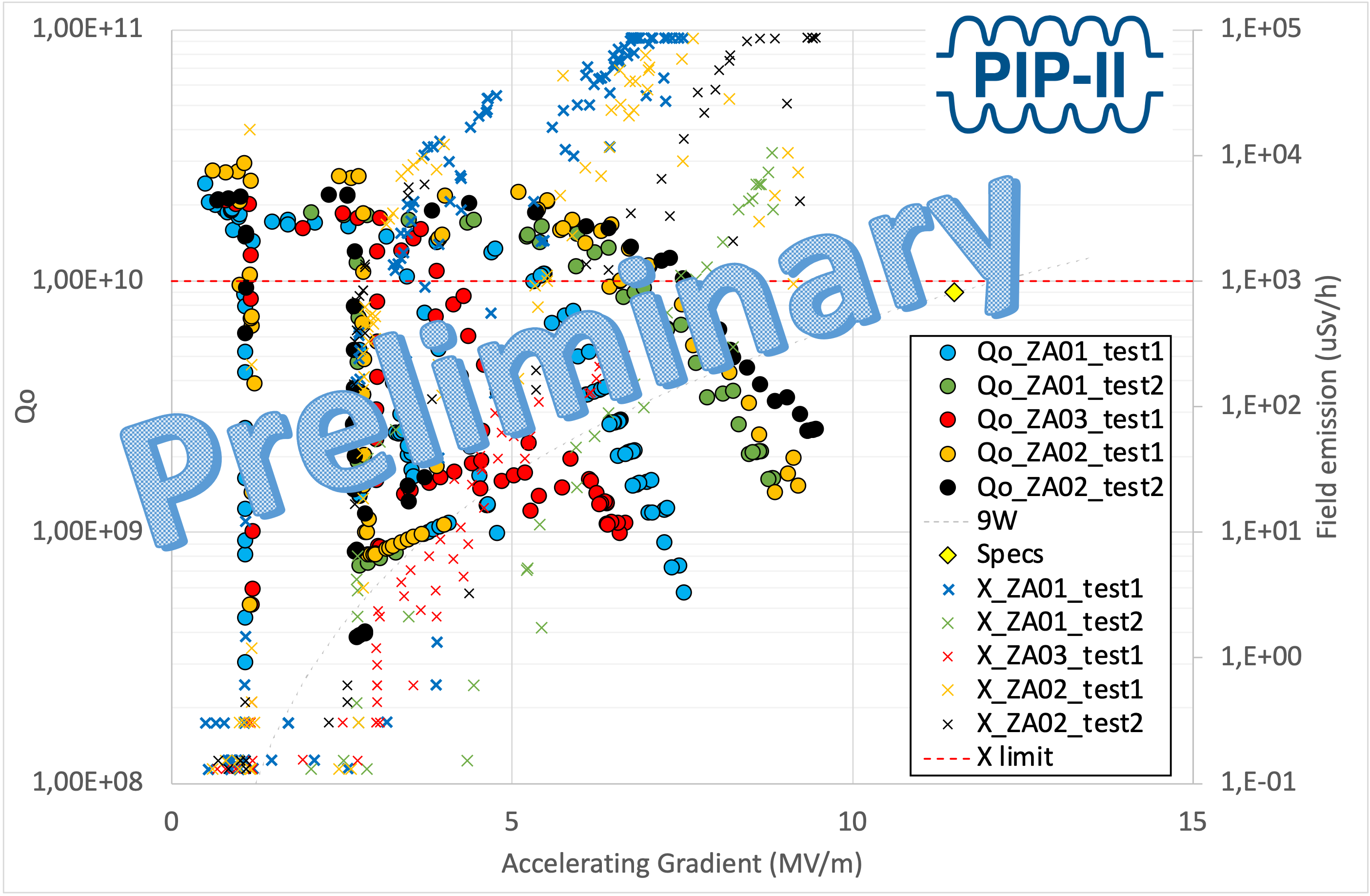}
   \caption{Preliminary results of PIPII prototype cavities\cite{PIPII_SRF2023_poster}.}
   \label{fig:PIPII}
\end{figure}

In the prototype SSR2, the cavity structure was designed to avoid MP bands at the nominal field level at the expense of a slight degradation of peak field ratios~\cite{Berrutti:2019tge}, 
being inspired by the balloon spoke cavities developed in TRIUMF~\cite{ZYYao2013StudyOB}.
IJCLab has tested three prototypes of this design with preliminary results shown in Fig.~\ref{fig:PIPII}~\cite{PIPII_SRF2023_poster}.
Surprisingly, we observed a deterministic field emission whose onset has been systematically around 4-5~MV/m in all the prototype cavities.
Moreover, MP bands at the low fields are substantially more difficult to condition compared to the ones in ESS and MYRRHA cavities.

The cause of the field emission is traced back to the fact that the designed SSR2 shape, optimized to avoid MP in the nominal fields, is so different from previous cavities that existing HPR tooling cannot cover the whole surface of the cavity.
During the design phase, a better RF performance was prioritized.
Fermilab and IJCLab are currently optimizing the HPR tooling to completely clean the complex surface of this spoke cavity.
We are forced to spend a factor of three times longer to pass the MP bands at low field.
Although the MP bands were well predicted at low field during design phase,
the strength and conditioning dynamic of MP is not predictable with the today code. 
Moreover, we are starting R\&D of preventive plasma processing during surface preparation that will definitely improve and speed up MP conditioning.

These preliminary results and discussions may imply important trade-offs among RF performance, surface cleaning, cold tests and operation for successful implementation of new cavities of complicated shape.
For example, as mentioned in this paper, the ESS series cavities are equipped with additional ports for HPR.
These ports enable horizontal and vertical BCP and therefore they offer thorough surface cleaning as well as fine tuning of the cavity frequency.
However, the ultimate reach of RF performance is slightly degraded by such ports, influencing the geometrical factors slightly.
The MP bands around the operational gradient are certainly of great concern but they have been easily conditioned (within 30 minutes) in the cold tests at IJCLab.
On the contrary, lower field MP needs several hours to overcome, even by experts of cavity measurement.
We could optimise RF performance but we might lose something else as a side effect.
These are important research subjects in the next years about global optimization in the spoke cavity technology.

\section{CONCLUSION}
IJCLab successfully tuned the frequency shifts of 29 series double-spoke cavties for the ESS project.
All the challenges in fabrication and processing were identified and were all solved so that the final frequency tolerance met the specification.
The ESS cavity performance was sufficiently beyond the project's specification.
All the ESS series cavities were assembled into cryomodules and passed the site acceptance test at Uppsala University and are being installed in the ESS tunnel.
The prototype single-spoke cavities for the MYRRHA project also showed very promising performance and the pre-series cavities are being fabricated by industry.
The new challenge is to industrialize chemical processing, heat treatment and HPR, following the recent success in the LCLS-II project for 1.3~GHz elliptical cavities.
The PIP-II SSR2 cavities are still in the prototyping phase.
Similar to the MYRRHA cavities, IJCLab plays a leading role in industrializing all the processes of cavity preparation.
One major difference from ESS and MYRRHA is its optimised RF design to avoid the MP bands at the nominal fields.
The MP bands is known to be problematic in spoke cavities.
Although this design challenge revealed another issue concerning surface cleaning,
we are optimising the cleaning process for this new geometry and will give feedback to the industry.
Another research subject is on baking spoke cavities and we pave the way to their operation at 4~K.

\section{acknowledgment}
We greatly appreciate the invaluable contributions from the FREIA laboratory during the series production of ESS spoke-cavity cryomodules.
We would like to acknowledge with appreciation the crucial role of colleagues from ESS.
We are deeply grateful to SCK-SEN and Fermilab for their leadership and cooperation in the MYRRHA and PIPII projects, respectively.
Last but not least, we thank all the technical staff, administrative colleagues, and in particular students, without whom the project and R\&D would have not and would not be feasible at all.

%
%
\ifboolexpr{bool{jacowbiblatex}}%
	{\printbibliography}%
	{%
	

} 
%
%


\end{document}